\documentclass[journal, twoside, onecolumn, draftclsnofoot,12pt]{IEEEtran}

\usepackage{multirow}  \usepackage{cite}
\usepackage{graphicx,subfigure}\usepackage{amsmath} \usepackage{amssymb} \usepackage{amsthm}\usepackage{color,array}
\usepackage{etex,etoolbox} \usepackage{float}
\usepackage{color}\usepackage{graphicx}
\usepackage{caption}\usepackage{xcolor}
\usepackage{colortbl}
\usepackage{soul}
\usepackage{algorithm}
\usepackage{soul,color,bm}
\usepackage{setspace}

\usepackage{algorithmic}   \usepackage{bbm}
\usepackage[T1]{fontenc}

\makeatletter
\providecommand{\@fourthoffour}[4]{#4}
\def\fixstatement#1{%
  \AtEndEnvironment{#1}{%
    \xdef\pat@label{\expandafter\expandafter\expandafter
      \@fourthoffour\csname#1\endcsname\space\@currentlabel}}}

\globtoksblk\prooftoks{1000}
\newcounter{proofcount}
\stepcounter{proofcount}
\long\def\proofatend#1\endproofatend{%
  \edef\next{ \Alph{proofcount}. Proof of \pat@label \noexpand\begin{proof}[Proof]}%
  \toks\numexpr\prooftoks+\value{proofcount}\relax=\expandafter{\next#1\end{proof}}
  \stepcounter{proofcount}}

\def\printproofs{%
  \count@=\z@
  \loop
    \the\toks\numexpr\prooftoks+\count@\relax
    \ifnum\count@<\value{proofcount}%
    \advance\count@\@ne
  \repeat}

\makeatother

\fixstatement{thm}
\fixstatement{lem}

\begin{document}


\title{Building Encoder and Decoder with Deep Neural Networks: On the Way to Reality}
\author{Minhoe Kim, ~\IEEEmembership{Member,~IEEE}, Woonsup Lee, ~\IEEEmembership{Member,~IEEE}, \\ Jungmin Yoon and Ohyun Jo, ~\IEEEmembership{Member,~IEEE}\thanks{ M. Kim is with the Department of Communication
	Systems, EURECOM, 06410 Sophia-Antipolis, France. W. Lee is with the Department of Information and Communication
	Engineering, Gyeongsang National University, Tongyeong 53064, South Korea. J. Yoon is with the Department of Electrical and Computer Engineering and Institute of New Media and Communications, Seoul National University, Seoul, 08826, South Korea. And O. Jo is the Department of Computer Science, College of Electrical and Computer Engineering, Chungbuk National University, Cheongju, 28644, South Korea. This work has been submitted to the IEEE for possible publication. Copyright may be transferred without notice, after which this version may no longer be accessible.}}

\maketitle



\begin{abstract}
Deep learning has been a groundbreaking technology in various fields as well as in communications systems. In spite of
the notable advancements of deep neural network (DNN) based technologies in recent years, the high computational complexity has been a major
obstacle to apply DNN in practical communications systems
which require real-time operation.
In this sense, challenges regarding practical implementation must be addressed
before the proliferation of DNN-based intelligent communications becomes a reality. To the
best of the authors' knowledge, for the first time, this article presents an efficient learning architecture and
design strategies including link level verification through digital circuit implementations using hardware description
language (HDL) to mitigate this challenge and to deduce feasibility and potential of DNN for communications systems. In
particular, DNN is applied for an encoder and a decoder to enable flexible adaptation with respect to the system
environments without needing any domain specific information. Extensive investigations and interdisciplinary design
considerations including the DNN-based autoencoder structure, learning framework, and
low-complexity digital circuit implementations for real-time operation are taken
into account by the authors which ascertains the use of DNN-based
communications in practice.
\end{abstract}

%
%
%
%
%
%
%
%
%
%
%
%
%
%
%
%
%
%
%
%
%
%
%

\section{Introduction}

Recently, deep learning has gained tremendous attention from academia, industry, and even
non-academic areas for its breaking through performance.
The basic principle of deep learning is based on mimicking the operation of neuron in human brain such that the multiple
layers of neuron nodes are stacked up with non-linear activation function between each layer, so called a deep neural
network (DNN)\footnote{In this article, we refer DNN as a broad definition which includes various types of neural
network such as RNN, CNN, FCnet, etc.}.
Such a structure is capable of extracting high and low level features from the input data where the features can be learned directly from the unrefined data rather than manual hand-crafted engineering.
Deep learning can be structured in different forms to cope with various types of problems.
For instance, sequential data, e.g., natural language, can be dealt with recurrent neural network (RNN) because there exists the temporal correlation
in data. Also, convolutional neural network (CNN) is effective to extract spatial features, while fully connected network (FCnet) is appropriate for tasks where all input data should be jointly considered.
Since deep learning has shown the
significant performance improvement in image classification
and speech recognition task, its application has recently been
considered extensively in various fields, which include the
communications systems.


As deep learning is known to have very powerful classification
capability, authors in \cite{OShea2016b} utilized CNN to
classify modulation level. Also, using the universal
approximation capability of DNN, the transmit power control scheme for
communications systems was proposed in
\cite{Sun2017} and  \cite{Lee2018}.
One of the most notable applications of deep learning in communications
systems is a DNN-based codec i.e., encoder and decoder.
There have been several attempts to apply DNN in the design of
an encoder and a decoder \cite{OShea2017a,OShea2017,Kim2018,Kim2018a,Ye2018a,Kim2018b,Kim2018c}.
In particular, autoencoder structure, which is originally
devised to reconstruct the original data from the corrupted
data, was adopted in the development of an encoder and a decoder.
In \cite{OShea2017a}, autoencoder structure was proposed to encode and decode Hamming code without prior knowledge
by adopting fully-connected networks (FCnet).
Moreover, the enhanced structure of DNN-based encoder and decoder was developed for
orthogonal frequency division multiplexing (OFDM)
systems \cite{Kim2018a,Ye2018a} using FCnet as well and a
variation of autoencoder structure was applied to build the encoder and decoder of
sparse code multiple access (SCMA) systems \cite{Kim2018}.
To that extend, RNN was adopted to find capacity achieving codes such as convolutional
code and turbo code in \cite{Kim2018b} which was followed by a proposal of designing feedback code with DNN in
\cite{Kim2018c}.

These precedent researches collectively showed theoretical possibility of DNN-based
communications systems which are able to adapt its operation according to
surrounding environments, e.g., dynamic channel, mobility, power, etc.
Although the benefits of DNN-based communications systems
have been extensively examined through simulations, the operation of DNN-based
communications systems in practice, especially with respect to the actual system architectures
and hardware implementation which is capable of real-time operation has
not been verified yet. Note that the realization of DNN-based
communications is not obvious because the DNN-based scheme can
require a large number of matrix operations. Thus
all major technical components constituting the DNN-based communications should interdisciplinary be studied to
overcome the challenges and to fully enable the artificial intelligence in
communications systems.

In this article, we describe major components which constitute deep
learning based communications down to the practical digital circuit designs including implementation and experimental results. Firstly, the comprehensible
explanation of deep learning structures is addressed to deliver the
principles of DNN-based autoencoder. We then introduce the learning framework that will be
applied for drastic reduction of learning overhead and hardware complexity of the DNN-based endpoint
devices as well as the details of our proposed DNN-based autoencoder which is flexible in terms of the
number of layers and the number of hidden nodes. The next section presents the in-depth description of the practical
design and considerations of the HDL-implementations. The effectiveness and accuracy of the DNN-based method are investigated and evaluated at the link level through the experiments followed up by conclusions.

\begin{figure*}[t!]
	\centerline{\includegraphics[width=1.1\textwidth]{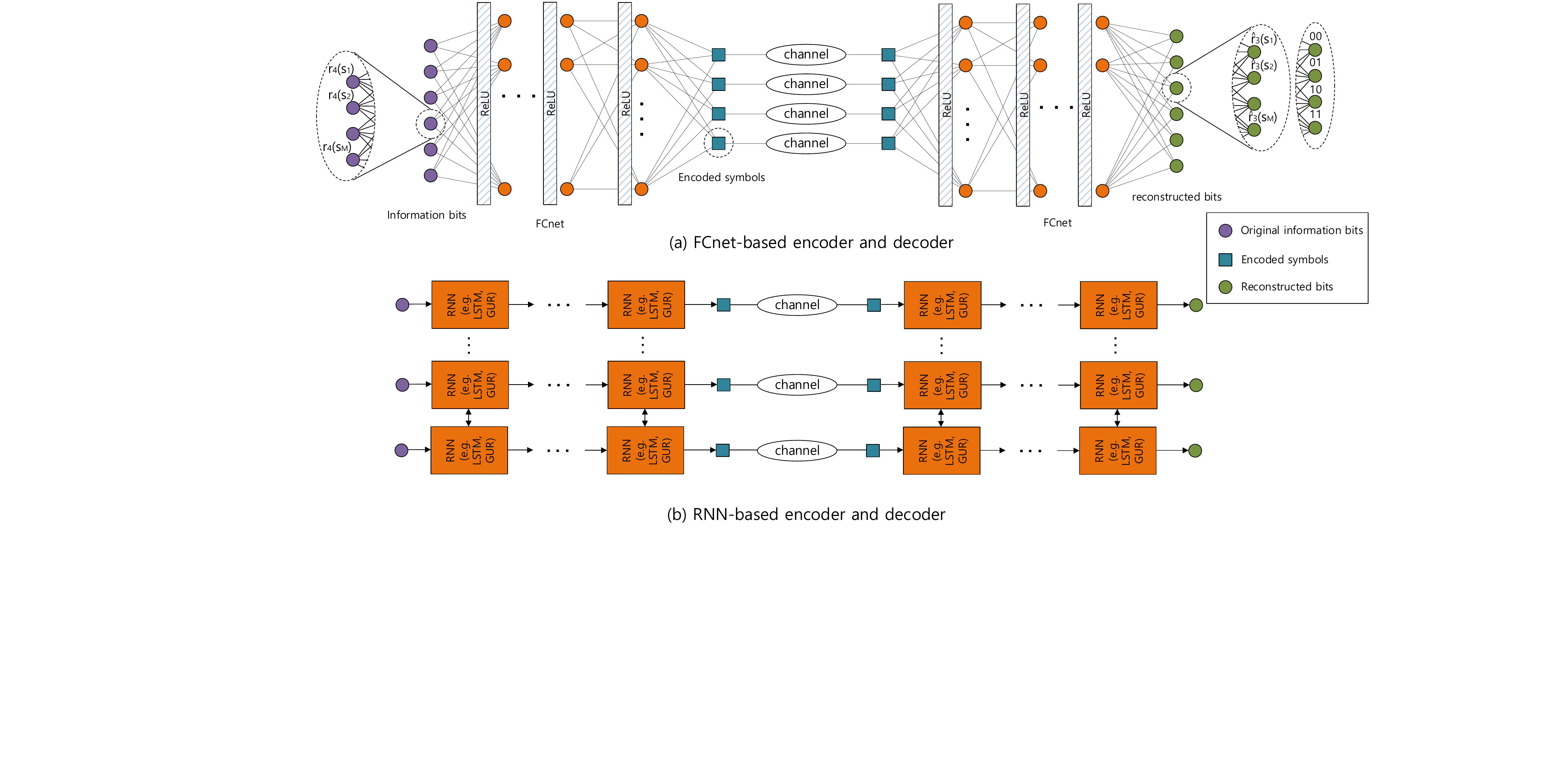}}
	\caption{Different types of DNN-based encoder and decoder.}
	\label{DNNautoencoder}
\end{figure*}

\section{Autoencoder for Communications}

In the DNN-based approach, an encoder is autonomously
found with DNN where input of the encoder is the information data, $r_n(s_m)$,
a decoder is also derived with DNN where the output of the decoder is the
reconstructed data, $\hat{r}_n(s_m)$, which should be the same with the input of the encoder.
Here, $r_n$ denotes the $n$-th information symbol and $s_m$ denotes $m$-th modulation bits as depicted
in Fig.
\ref{DNNautoencoder}.
Therefore, the goal of the DNN encoder and decoder is to minimize the loss between original data and reconstructed date
which comes down to the loss function $\mathcal{L}(r_n, \hat{r}_n)$, where the loss function $\mathcal{L}$ can be any
distance
function, e.g., $L_2$ norm or cross entropy.
The output of the encoder can be generally expressed as $f(\bm{r};\theta_f)$ and the output of the
decoder is
$\bm{\hat{r}} = g(H(f(\bm{r};\theta_f))+\bm{\epsilon};\theta_g)$ where $\theta_f$ and $\theta_g$ refer
to parametric
representation of
the DNN for encoder and decoder respectively \cite{Kim2018}, $H(\cdot)$ denotes the channel gain
and $\bm{\epsilon}$ denotes
additive noise.
Thereby, the final goal of the DNN autoencoder can be expressed an optimization
problem whose objective is $\min\limits_{\theta_f, \theta_g} \mathcal{L}(\bm{r}, \bm{\hat{r}}) = \mathcal{L}(\bm{r} ,
g(H(f(\bm{r};\theta_f))+\bm{\epsilon};\theta_g))$.

One interesting characteristic of the DNN-based encoder and decoder is the flexible applicability.
The structure of $f(x;\theta_f)$ and $f(x;\theta_g)$ is not limited to FCnet or RNN but other types of DNN can be used instead as long as they are
able to encode or decode the signal by taking account whole data sequence $\bm{x}$.

Also, by changing the size of the output of encoder, the code rate can also be controlled. For example, if the size of
encoder output bits is twice the size of the original information bits, the code rate will be 1/2.
Moreover, when the encoded sequence is transmitted parallelly, the DNN can be applied at parallel transmission systems
such as OFDM, on the other hand, when the the transmitted data is encoded sequentially, the DNN can be applied as a
forward error correcting code.
In Fig. \ref{DNNautoencoder}, possible DNN architectures are presented that can be applied to an encoder and a decoder
of different systems.
FCnet designed for a multi-carrier wireless communications system is shown in Fig. \ref{DNNautoencoder}(a)
\cite{Kim2018a}. By using FCnet,
block information bits can be handled at the same time so that the parallel encoding is enabled. This structure is
suitable for OFDM system \cite{Ye2018, Kim2018a} which is taken account for HDL-implementation in this work.
Also, the structure of a bi-directional RNN-based encoder and decoder is shown in Fig.
\ref{DNNautoencoder}(b)
\cite{Kim2018b} where the basic building block is a single RNN, i.e., popular RNN types such as long short term memory
(LSTM) or gated recurrent unit (GRU) can be used.
In particular, the bi-directional RNN takes account of the neighbor information bits by feeding in the signals from
neighbor RNN blocks. In other words, the encoded
codes are highly correlated with the previous and the next bits, which is similar to convolutional codes.
These structures have the flexibility to control the code rate by changing the ratio between
the input information bits and the output bits.

In perspective of computational complexity, building an encoder and a decoder with DNN may be challenging because it
requires large amount of arithmetic computations, i.e., multiplications and additions, for matrix operations in DNN.
Therefore, in order to accelerate operations in the DNN-based encoder and decoder, it is inevitable to implement a dedicated and
pipelined digital circuit designs in which the processing time can be reduced to support the high data rate.

\begin{figure*}[t]
	\centerline{\includegraphics[width=1.1\textwidth]{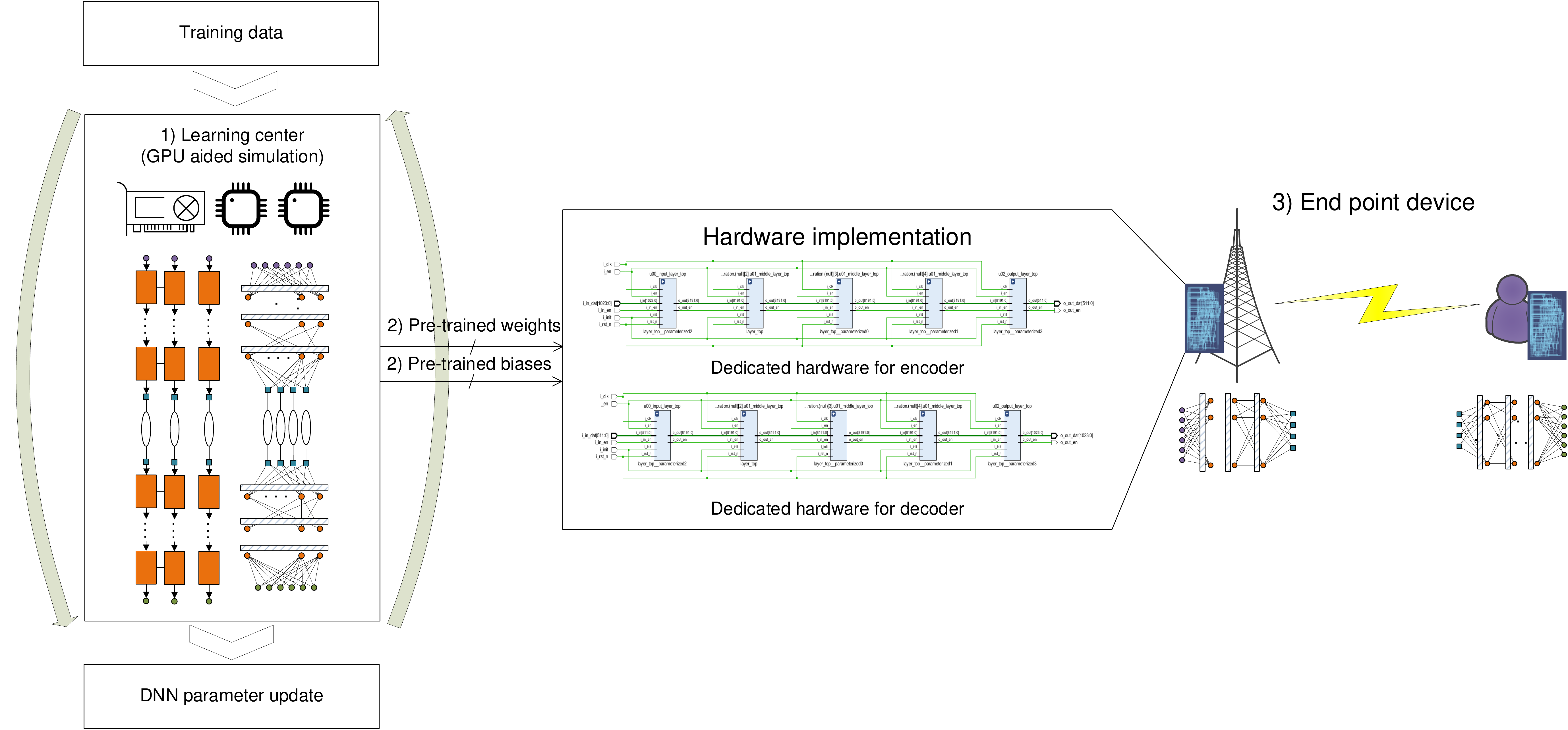}}
	\caption{Proposed learning framework for training and implementing DNN-based encoder and decoder.}
	\label{Training}\vspace{-5mm}
\end{figure*}

\section{Learning Framework for End-to-end Communications Systems}

In this section, a learning framework for end-to-end communications systems is described in
view of practical operation for the communications systems adopting the DNN-based encoder and decoder.
As many people concerns, the training process of deep learning requires a lot of computations and time to sufficiently
train a properly working DNN.
The training can be possibly accelerated with the aid of graphics processing unit (GPU), but still, the
training process cannot be done in the real-time operation.
On the other hand, if the pre-trained DNN is available, the inference process can be done in the real-time operation.
Therefore, the training process and the inference process are separated at the learning center
with large computation power and at the endpoint device with high-speed
and real-time operational digital circuits, respectively.
The aforementioned learning framework is depicted in Fig. \ref{Training} and the major components of the learning framework are described as follows.

\subsection{Components of Learning Framework}

\subsubsection{Learning center}
In the learning center, the DNN parameters of the DNN-based encoder and decoder are learned with the
actual channel
data or synthesized channel data if actual channel data is insufficient\footnote{When the synthesized channel
data are used, DNN can be initially trained with synthesized channel data and then the DNN
	parameters can be adjusted base on the actual data where such a scheme is so called as \textit{transfer learning}
	\cite{pan2010survey}.}.
This learning center is necessarily equipped with parallel computing devices, e.g., GPU, so that the DNN parameters can
be learned efficiently and computation time can be reduced.
It should be noted that the DNN structure trained in the learning center must be the same DNN structure which is
implemented on the modem chip so that only the pre-trained parameter information is delivered.

\subsubsection{Parameter delivery}
After sufficient training, the DNN parameters for an encoder and a decoder are supposed to
be delivered to endpoint communications devices, i.e., base station (BS) and user equipment (UE), or can be set at the
initial factory production phase in the proposed learning framework.
The DNN parameters, such as weights and biases, which are pre-determined by the learning center, are
put into the encoder and
decoder of each device.

\subsubsection{Endpoint device}
For endpoint devices, the encoder and decoder modules are built with dedicated hardware for boosting the signal
processing where the weights and the biases are fed in by the learning center.
The DNN-based encoder and decoder have been already informed with the pre-trained
DNN parameters, thereby they only require one-shot inference process for one transmission.
The DNN parameters can be updated periodically or whenever the better DNN parameters are learned or
updated adaptively when channel environment has changed.

\subsection{Overhead Reduction in the Proposed Framework}
The proposed framework and the training process are beneficial because the training
process does not take place at the endpoint device. Therefore, the overhead originated from the
high hardware cost and huge power consumption of the training process can be effectively removed so that the computing
unit, e.g. GPU, can be excluded from endpoint devices.
Also, given that the encoder and decoder are built in the form of dedicated hardware which requires only the inference computation,
the real-time operation is supportable and the power consumption is greatly reduced at endpoint devices where the power
consumption is of critical issue.


\begin{figure*}
	\centerline{\includegraphics[width=1.1\textwidth]{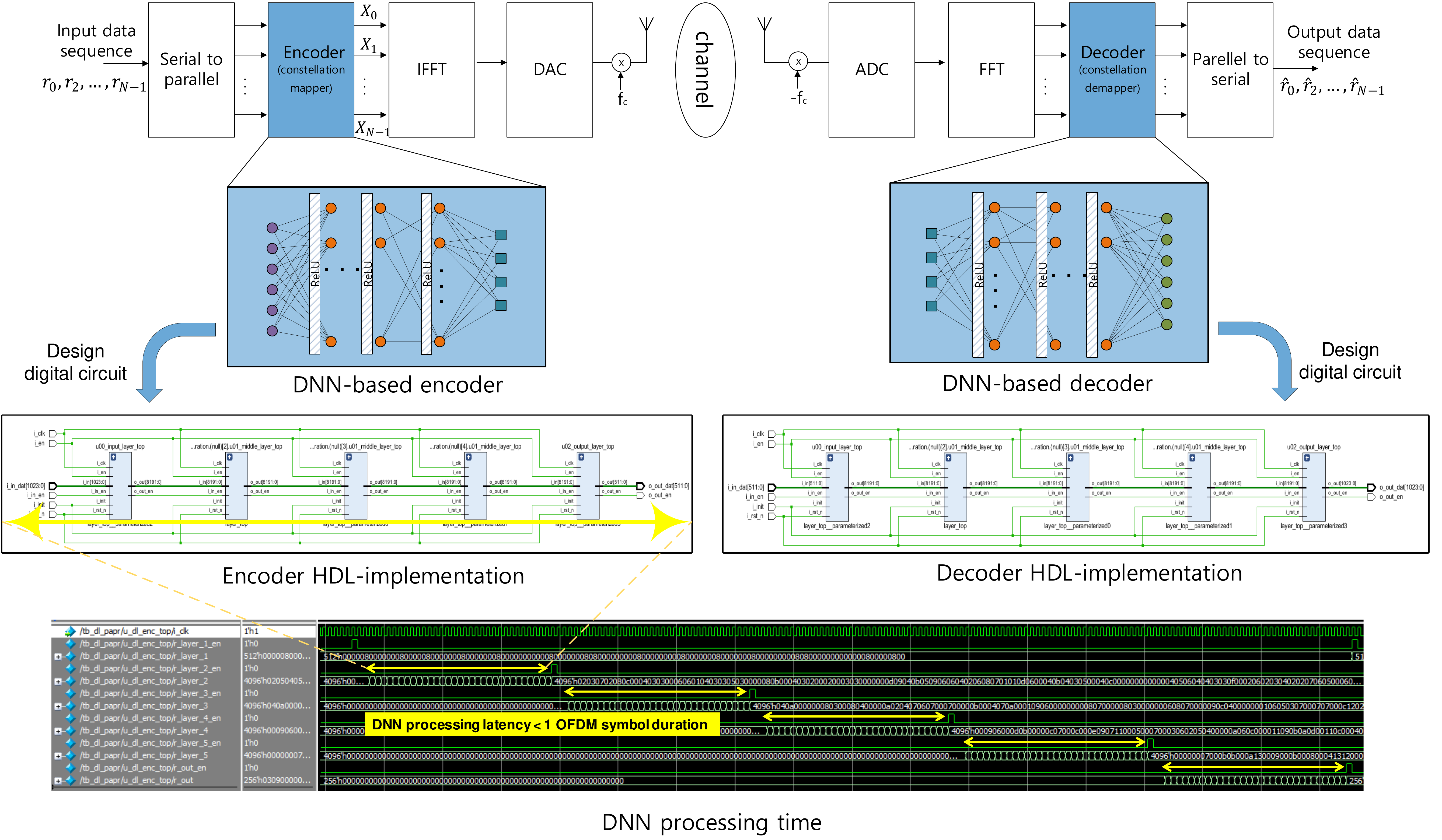}}\vspace{-1mm}
	\caption{Implementation for DNN-based encoder and decoder of OFDM system.}
	\label{RFchain}\vspace{-5mm}
\end{figure*}

\section{Hardware Implementation of Real-time DNN-based Encoder and Eecoder by Using HDL}

In this section, we introduce design strategies and experiences for implementation of HDL-based
hardware digital
circuits which can improve competitiveness and feasibility of the DNN-based communications systems. In reality,
signal processing for data transmission and reception in practical communications systems and devices should be carried
out immediately and accurately to support the high data rate. Even though the DNN-based encoder and decoder can have
benefits in terms of flexibility and usability, they are usually considered to require high computational complexity
and large amount of computing resource compared to legacy systems. A huge amount of training data further exacerbates
the difficulty of the practical implementation of the DNN-based communications technologies.
Consequently, deep learning structure and the training process should reflect the system requirements. Related to this
fact, we have
discussed in section III about the learning framework and the parameter delivery procedure which can be applicable in
general. Based on this framework, we present the design strategies for realizing the DNN-based communications systems
as well as our hands-on experience and lessons.


\subsection{HDL-based DNN Encoding and Decoding}
As is known, in perspective of system design, there always exists tradeoff between software and hardware: flexibility
in implementation vs. promptness in processing \cite{jo2014holistic}. This is also applied for implementation of
DNN. Thus, as shown in Fig. \ref{RFchain}, we devised a novel DNN system architecture for OFDM system
where the encoder and decoder is based on DNN and implemented it
by HDL to show the practical feasibility of the deep learning-based communications systems while focusing on the DNN-based encoder
and decoder. This work includes the well-collaborating functionalities of the learning framework and hardware implementation. As stated early,
fast and accurate data transmission is the most significant attribution of the recent communications systems.
In our DNN structure, the essential functionalities of DNN are included in the
shape of digital circuits for boosting the signal processing and for obtaining the advantages of hardware
implementation.



\subsection{Hardware Complexity}

Generally, computational complexity is highly related to the size and power consumption of
the hardware, i.e., digital modem integrated circuit (IC). Thus, the hardware structure of fully-connected sublayers,
which have high complexity,
should be efficiently devised. When the number of hidden nodes in a fully-connected sublayer is $L$,
the computational complexity increases in the order of $O(L^2)$. The functional structure of the fully-connected sublayer of the DNN
structure is similar to the $L$-point fast fourier transform (FFT) block which is widely used and efficiently
implemented in the practical OFDM communications systems. However,
effective computational complexity of the fully-connected sublayer is higher than the FFT because the well-known
butterfly algorithm can reduce the complexity of the FFT in the order of $O(L log(L))$. In this respect, the DNN
structure should have the optimal number of hidden node for preventing too many computations, while guaranteeing the required target performance. On the other hands, the number of layers
also affects the complexity of the DNN implementation. The computational complexity of DNN
proportionally increases according to the number of the layers. Bit width is also important because the representation of signal value should be a fixed point format in hardware implementation. Moreover, the implementation loss can be further accumulated as the
number of layers increases. Thus, it is required to optimize the number of layers, and the number of hidden nodes and the bit width for data representation.

\subsection{DNN Requirement for Real-time Operation}

Processing latency is one of the most critical requirements for enabling real-time operation of DNN-based
communications. The total processing time across all the components consisting of transmission link (including serial to parallel
module, modulator, encoder, IFFT, and others) and reception link (including FFT, decoder, demodulator, parallel to serial module, and others)
should be able to support the data rate. For example, the total
processing latency for transmission link and reception link in Fig. \ref{RFchain} should be less than the time duration of
an OFDM symbol. In the respect of HDL design, efficient structures for pipelining and parallel computation, system
clock divider should be devised for practical implementation to satisfy the tight requirement due to the high
complexity of DNN.

\subsection{HDL-implementation}

\begin{figure*}[t]
	\centerline{\includegraphics[width=\textwidth]{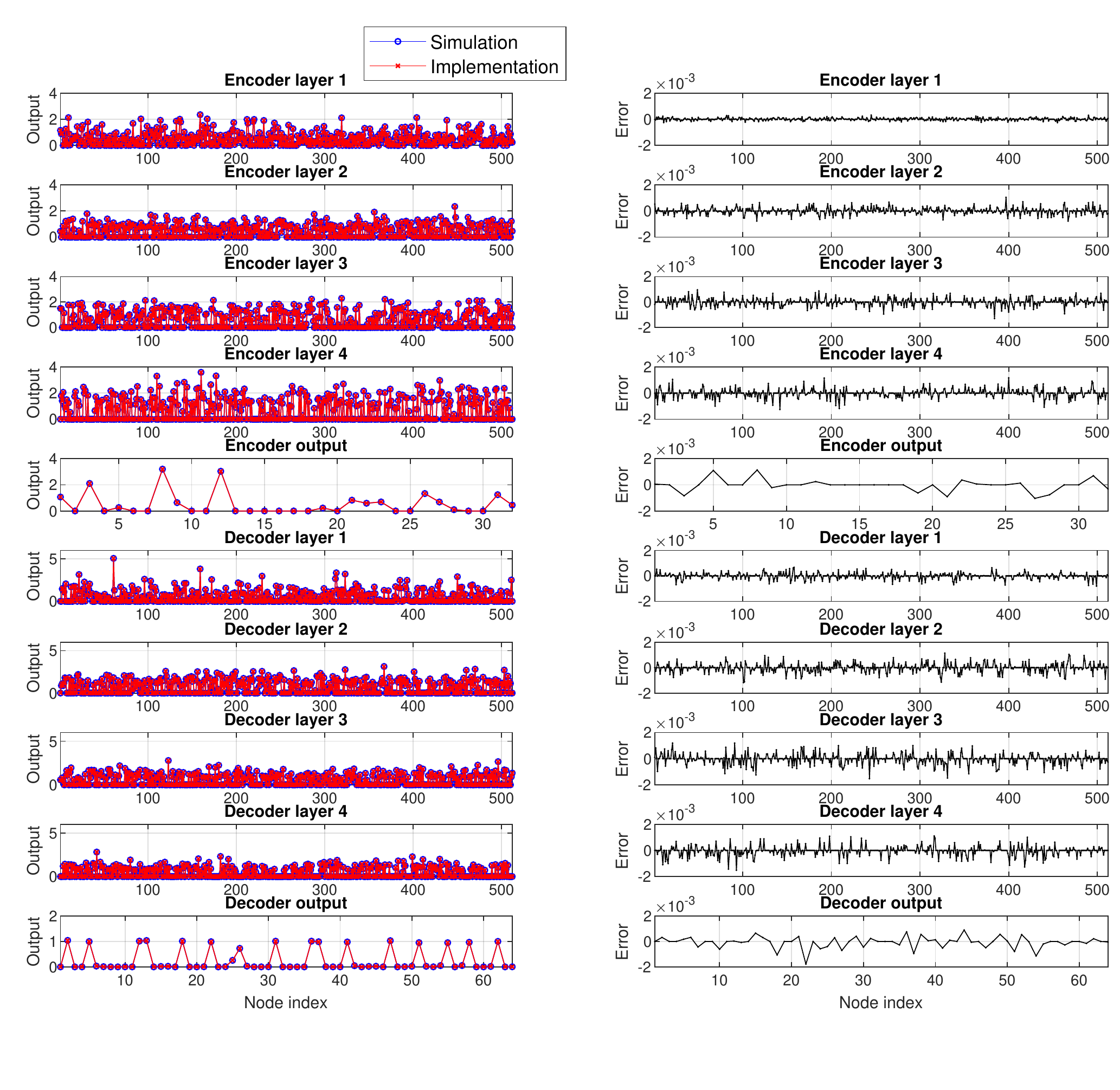}}\vspace{-1mm}
	\caption{Measurements of signal and error from HDL design and software simulation of each layer.}
	\label{overall}\vspace{-5mm}
\end{figure*}

To verify the practical feasibility, we have implemented hardware digital logic of the proposed DNN by using Verilog and confirmed the correct operation and performance by using ModelSim, a multi-language HDL design utility. Link level experiments are performed under the proposed end-to-end learning framework and learning scenarios depicted in Fig. \ref{Training}. Thus, empirical investigations which are taken under practical hardware and software environments ascertain that the HDL-based DNN can alleviate complexity
of the deep learning for real-time communications. Representatively, the processing time of each DNN layer is less than 1 OFDM symbol duration as shown in the bottom of Fig. \ref{RFchain}. And the error vector magnitude under the DNN configuration constituting 5 layers with 512 hidden nodes is almost negligible as shown in the Fig. \ref{overall}. Also, DNN-based link level test achieved the minimal bit error rate (BER) which is less than $10^{-3}$ at 30 dB SNR in rayleigh channel.

\begin{figure}
	\centerline{\includegraphics[width=0.8\textwidth]{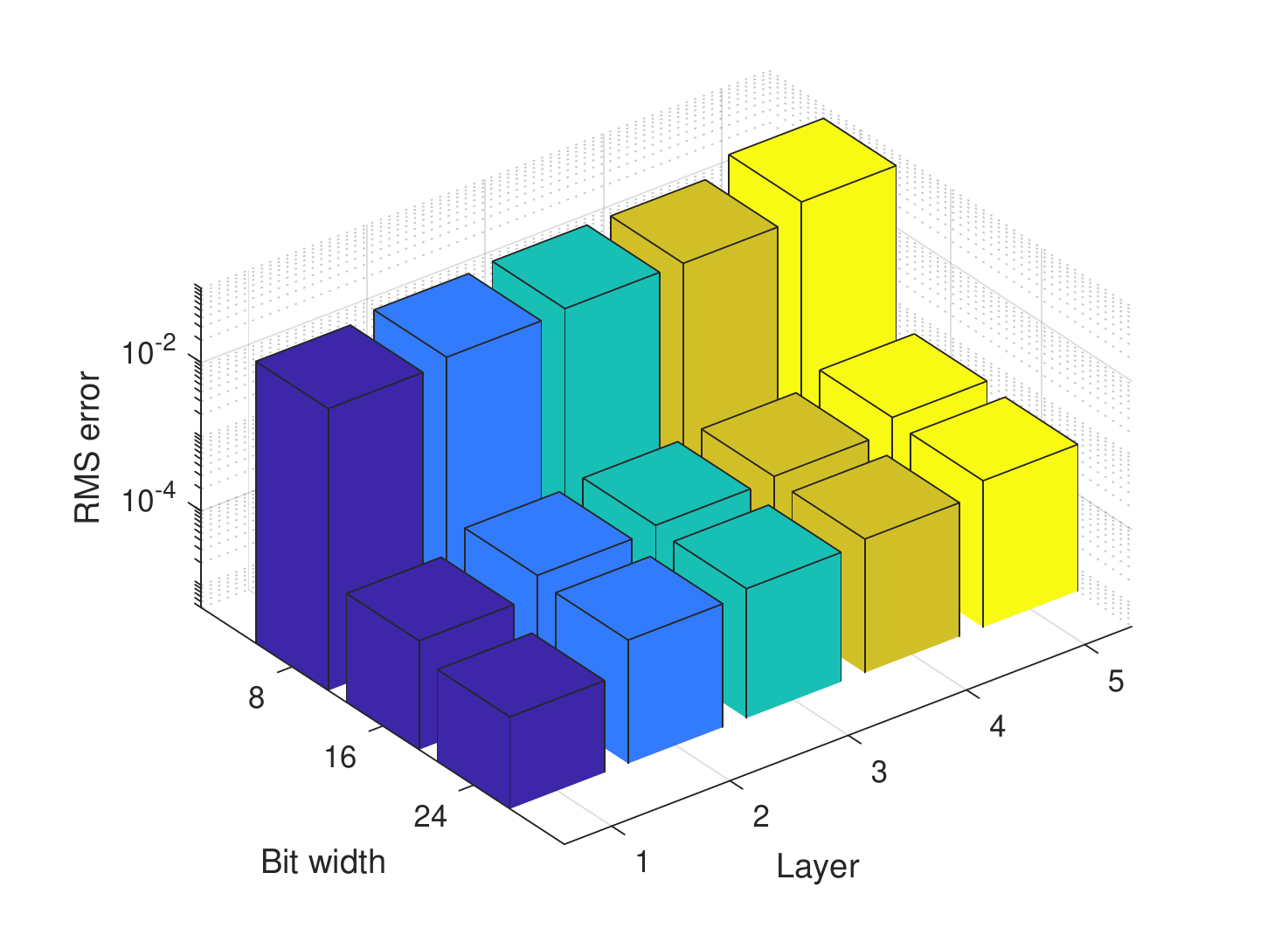}}\vspace{-1mm}
	\caption{RMS error propagation through DNN layers for different bit width.}
	\label{bitwidth}\vspace{-5mm}
\end{figure}
\begin{figure*}
	\centering
	\subfigure[RMS error of different hidden noded DNN structures (5 layers).]{
		\centering
		\includegraphics[width=0.5\textwidth]{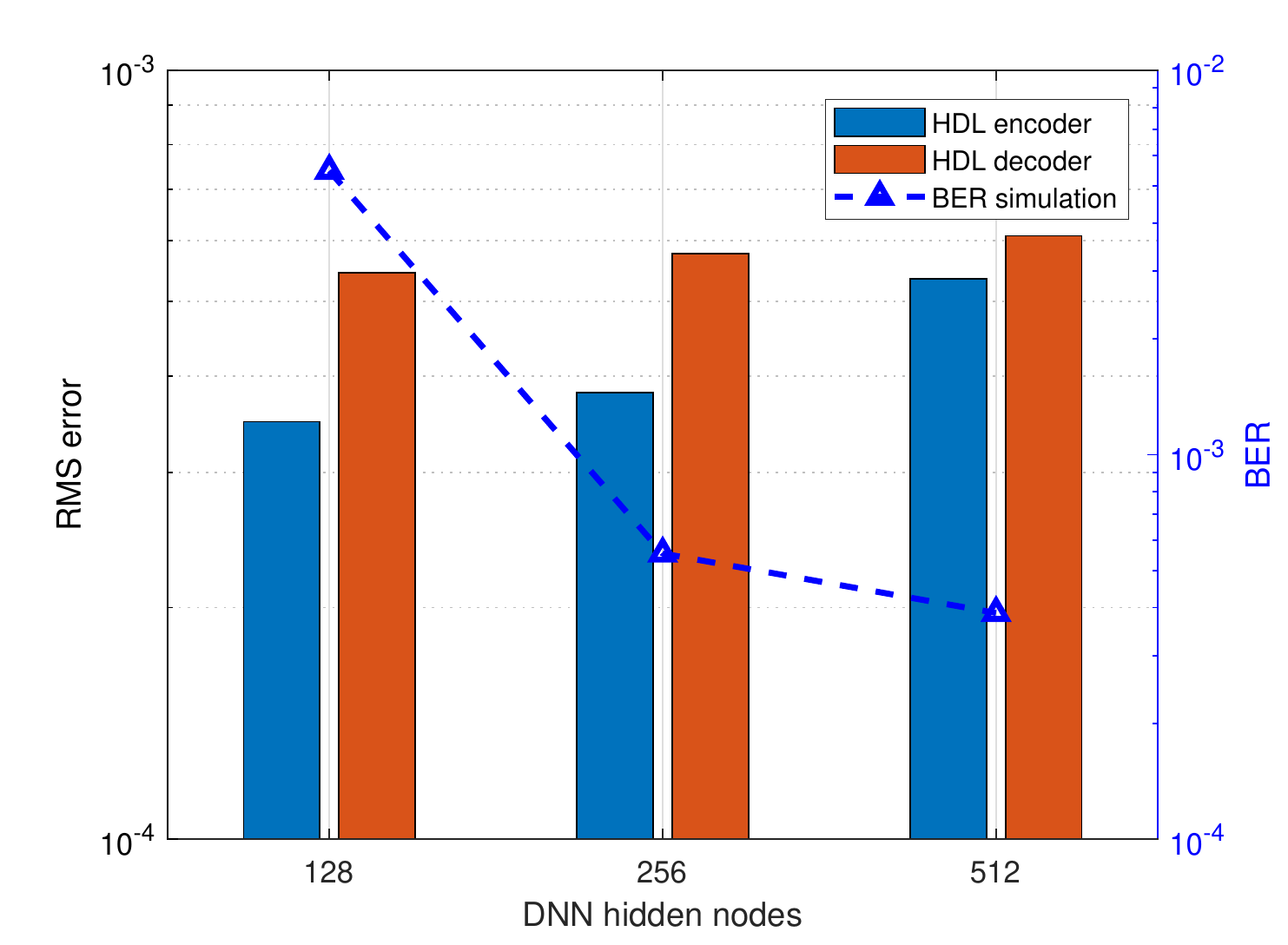}
	}%
	\subfigure[RMS error of different layered DNN structures (512 hidden nodes).]{
		\centering
		\includegraphics[width=0.5\textwidth]{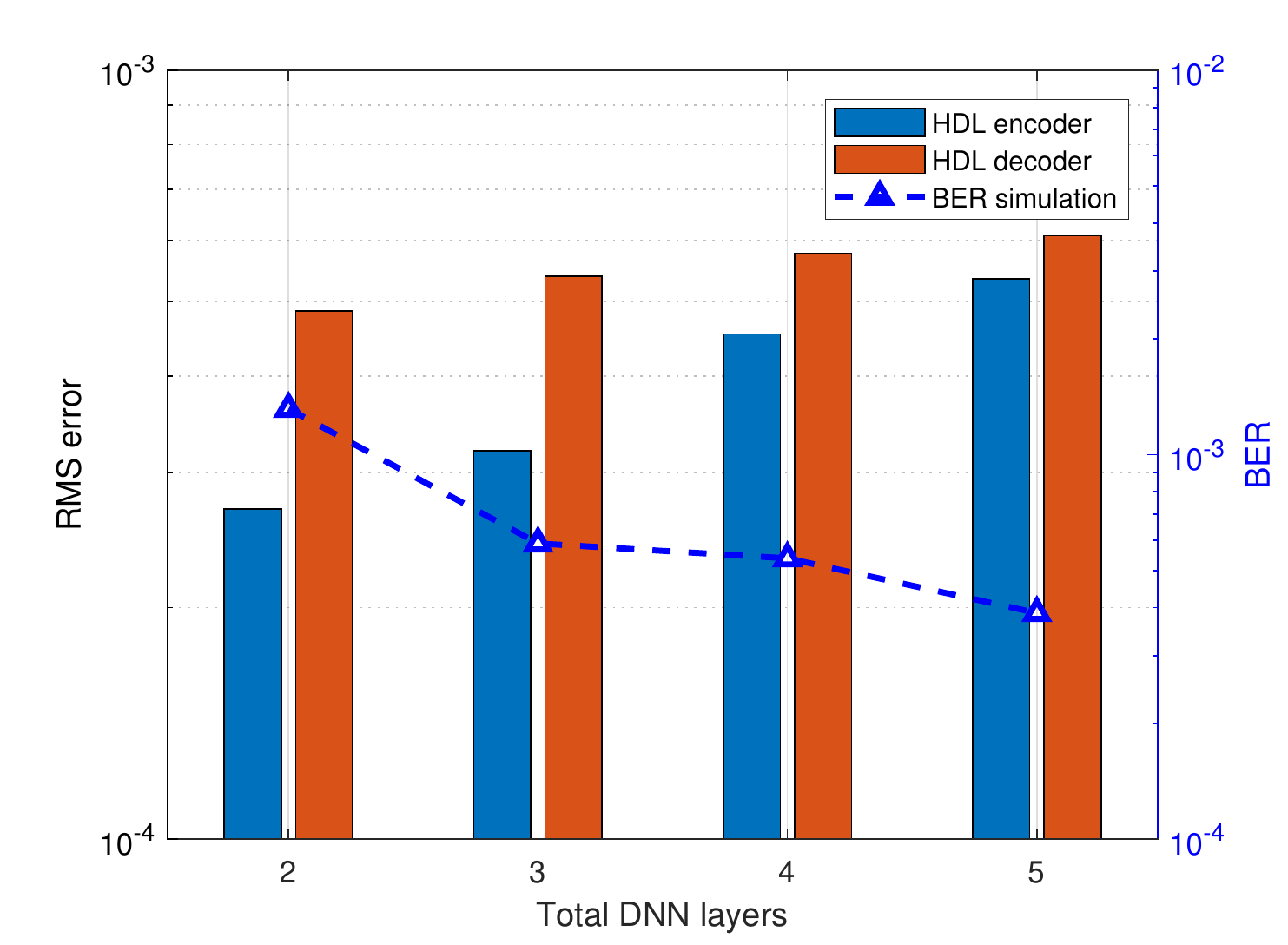}
	}
	\caption{RMS error between HDL design and simulation for different structures of DNN-based encoders and decoders}
	\label{RMSerror}\vspace{-5mm}
\end{figure*}

\subsection{Experimental Results and Lessons}

Fig. \ref{overall} describes the test results across the entire communications transmission and
reception links based on HDL-implementations. OFDM system of 16 subcarriers and 4-QAM modulation is taken account for implementation.
The righthand side of Fig. \ref{overall} shows the results of floating point simulation using MATLAB and fixed point HDL implementation using Verilog.
The DNN structure devised for the encoder and decoder in this work may have variable number of layers and hidden nodes, but the test results in Fig. \ref{overall} are obtained in the case of 5 layers and 512 hidden nodes. In each layer, we compared the output value of simulation with the output value of HDL-implementation to confirm the accuracy and the correct functionality of the DNN hardware digital circuit. For all layers and all hidden nodes composing the encoder and the decoder, the HDL-implementation have same tendency with the simulation. Here, the horizontal axis means the index of the hidden node in each layer. The lefthand side of Fig. \ref{overall} shows the results of HDL-implementation error, which is exactly the gap between the output value of simulation and the output value of the HDL-implementation in the righthand side of Fig. \ref{overall}. The amount of error due to the HDL-implementation loss is measured and confirmed to be within the acceptable bound
while satisfying the hardware requirements regrading total processing latency and computational complexity. Note that the 64-bit floating point is used in and 16-bit fixed point is used in HDL-implementations in the test results shown in Fig. \ref{overall}. The amount of implementation error is less than 0.2\%.

Interestingly, it can be seen that the implementation loss is propagated further as the signal passes DNN
layers one by one. Fig. \ref{bitwidth} summarizes the results regarding the
aforementioned error propagation. According to the axis of layer, the root mean square (RMS) error increases for all
cases. That means, the individual error components in front layers are accumulated at the last layer of the DNN. It
tells that the number of layers may affect the system performance from the perspective
of HDL-implementation, even if the recent deep learning networks such as GoogleNet and Residual Net \cite{Szegedy2015}
showed that the ideal performance of DNNs can be improved by adopting very huge number of hidden layers. We can realize that the number of layers can not be too large when it comes to hardware implementation.

Fig. \ref{bitwidth} also shows that the system performance is highly
affected by the data bit width in HDL-implementation. Herein, the data bit width means the width of data bus in the
implemented DNN hardware, which is in shape of digital circuit. For representing the fixed-point format, one bit is
allocated for sign, 3 bits are allocated for integer part, and the other bits are allocated for fractional part. Then, we
observed 3 cases of bit width, which are 8-bit, 16-bit, and 24-bit. As we expected, when the small bit width is used the RMS
error is drastically increased. For example, the RMS error is more than 100 times in case of 8 data bits, compared to
the case of 16 data bits. If too small data bit width is used in practical HDL-implementation, the system performance
is degraded significantly even if the computational overhead can be reduced. We can see that the data bit width does
not need to be too large. That is to say, in the case of 24 data bits, the RMS error is not improved impressively even if the
computational overhead increase drastically.

Fig. \ref{RMSerror} shows the tendency of RMS error and bit error rate (BER) according to the total number of hidden
nodes in each layer and the number of hidden layers of the DNN, respectively. Obviously, as shown in Fig.
\ref{RMSerror}(a), the number of hidden nodes in each DNN layer is also a very important design parameter which affects
the system performance. The large number of hidden nodes may improve the BER performance of the DNN itself as
shown in the dashed line in Fig. \ref{RMSerror}(a), however, in the algorithmic view, it is hard to train DNN
parameters in case of large number of hidden nodes due to the vanishing gradient problem. From the perspective of
practical implementation, large number of hidden nodes may increase the number of possible error sources and accumulate
larger error further as shown from bar graphs in Fig. \ref{RMSerror}(a). Thus, larger number of hidden nodes does not
always guarantee better performance in hardware implementation perspective.

The number of layers is another critical design factor in DNN. It is represented as the depth of the DNN in other
words. The large number of layers in DNN may improve the BER performance of the DNN itself as shown in the dashed line
in Fig. \ref{RMSerror}(b). The BER performance improved more than 10 times in case of 5-layer DNN compared to the case
of 2-layer DNN. However, from the perspective of practical implementation, large number of layers increases RMS error
as shown from bar graphs in Fig. \ref{RMSerror}(b). This is because the error components are propagated and accumulated
further in more deeper learning networks. Thus, the these design parameters should be collectively considered to
optimize the DNN-based communications.

We have shown examinations of the DNN-based encoder and decoder in view of the bit width, the number of
hidden nodes, and the hidden layers.
From the results, we can see that the increasing the bit width can reduce the error between software simulation
and hardware design, but does not improve the error performance effectively after a certain number of bit width.
Since the bit width affects the complexity of the hardware, the number of bit width should be chosen carefully for
effective error reduction.
Also, we have witnessed the trade-off between the hardware-implementation error and BER performance with respect to the number
of hidden nodes and the number of hidden layers.
Increasing the number of hidden nodes or the hidden layers of DNN can increase the BER performance, but it also increases
the hardware-implementation error.
Therefore, the number of hidden nodes and the number of hidden layers
should not be too large in case of hardware implementation and carefully determined regarding both the BER performance and the implementation error for optimizing system performance.

\section{Conclusions and Further works}

In the investigations on the HDL-based DNN communications systems, we
have been confirmed the feasibility of the DNN-based communication
technologies for enhancing the flexibility and productivity of the systems
through actual HDL-implementation. Especially, the observed measurement
has highlighted the applicability and potential of the proposed
DNN algorithms for communications systems in practice.
Future works remain in bridging the link level HDL-implementation and designed strategies with the system
level verification by using field programmable gate array (FGPA) or application specific integrated circuit (ASIC)
implementation for the maturity of innovative applications combining deep learning and communications.


\bibliographystyle{IEEEtran}
\bibliography{IEEEabrv,mybibfile}

\begin{thebibliography}{10}
\providecommand{\url}[1]{#1}
\csname url@samestyle\endcsname
\providecommand{\newblock}{\relax}
\providecommand{\bibinfo}[2]{#2}
\providecommand{\BIBentrySTDinterwordspacing}{\spaceskip=0pt\relax}
\providecommand{\BIBentryALTinterwordstretchfactor}{4}
\providecommand{\BIBentryALTinterwordspacing}{\spaceskip=\fontdimen2\font plus
\BIBentryALTinterwordstretchfactor\fontdimen3\font minus
  \fontdimen4\font\relax}
\providecommand{\BIBforeignlanguage}[2]{{%
\expandafter\ifx\csname l@#1\endcsname\relax
\typeout{** WARNING: IEEEtran.bst: No hyphenation pattern has been}%
\typeout{** loaded for the language `#1'. Using the pattern for}%
\typeout{** the default language instead.}%
\else
\language=\csname l@#1\endcsname
\fi
#2}}
\providecommand{\BIBdecl}{\relax}
\BIBdecl

\bibitem{OShea2016b}
T.~J. {O'Shea}, J.~Corgan, and T.~C. Clancy, ``Convolutional radio modulation
  recognition networks,'' in \emph{Proc. of EANN}, Aberdeen, U.K., Sep. 2016.

\bibitem{Sun2017}
H.~Sun, X.~Chen, Q.~Shi, M.~Hong, X.~Fu, and N.~D. Sidiropoulos, ``Learning to
  optimize: Training deep neural networks for wireless resource management,''
  \emph{arXiv preprint arXiv:1705.09412}, 2017.

\bibitem{Lee2018}
W.~Lee, M.~Kim, and D.~H. Cho, ``Deep power control: {T}ransmit power control
  scheme based on convolutional neural network,'' \emph{{IEEE} Commun. Lett.},
  vol.~22, no.~6, pp. 1276--1279, 2018.

\bibitem{OShea2017a}
T.~O'Shea and J.~Hoydis, ``An introduction to deep learning for the physical
  layer,'' \emph{IEEE Transactions on Cognitive Communications and Networking},
  vol.~3, no.~4, pp. 563--575, 2017.

\bibitem{OShea2017}
T.~J. O'Shea, T.~Erpek, and T.~C. Clancy, ``Deep learning based mimo
  communications,'' \emph{arXiv preprint arXiv:1707.07980}, 2017.

\bibitem{Kim2018}
M.~Kim, N.~I. Kim, W.~Lee, and D.~H. Cho, ``Deep learning aided {SCMA},''
  \emph{{IEEE} Commun. Lett.}, vol.~22, no.~4, pp. 720--723, Apr. 2018.

\bibitem{Kim2018a}
M.~Kim, W.~Lee, and D.~H. Cho, ``A novel {PAPR} reduction scheme for ofdm
  system based on deep learning,'' \emph{{IEEE} Commun. Lett.}, vol.~22, no.~3,
  pp. 510--513, Mar. 2018.

\bibitem{Ye2018a}
H.~Ye, G.~Y. Li, and B.-H. Juang, ``Power of deep learning for channel
  estimation and signal detection in ofdm systems,'' \emph{IEEE Wireless
  Communications Letters}, vol.~7, no.~1, pp. 114--117, 2018.

\bibitem{Kim2018b}
H.~Kim, Y.~Jiang, R.~Rana, S.~Kannan, S.~Oh, and P.~Viswanath, ``Communication
  algorithms via deep learning,'' \emph{arXiv preprint arXiv:1805.09317}, 2018.

\bibitem{Kim2018c}
H.~Kim, Y.~Jiang, S.~Kannan, S.~Oh, and P.~Viswanath, ``Deepcode: Feedback
  codes via deep learning,'' \emph{arXiv preprint arXiv:1807.00801}, 2018.

\bibitem{Ye2018}
H.~Ye, G.~Y. Li, and B.~H. Juang, ``Power of deep learning for channel
  estimation and signal detection in {OFDM} systems,'' vol.~7, no.~1, pp.
  114--117, Feb. 2018.

\bibitem{pan2010survey}
S.~J. Pan, Q.~Yang \emph{et~al.}, ``A survey on transfer learning,''
  \emph{{IEEE} Trans. Knowl. Data Eng.}, vol.~22, no.~10, pp. 1345--1359, Oct.
  2010.

\bibitem{jo2014holistic}
O.~Jo, W.~Hong, S.~T. Choi, S.~Chang, C.~Kweon, J.~Oh, and K.~Cheun, ``Holistic
  design considerations for environmentally adaptive {60 GHz} beamforming
  technology,'' \emph{{IEEE} Commun. Mag.}, vol.~52, no.~11, pp. 30--38, Nov.
  2014.

\bibitem{Szegedy2015}
C.~Szegedy, W.~Liu, Y.~Jia, P.~Sermanet, S.~Reed, D.~Anguelov, D.~Erhan,
  V.~Vanhoucke, and A.~Rabinovich, ``Going deeper with convolutions,'' in
  \emph{Proc. of IEEE CVPR}, Boston, MA, USA, Jun. 2015, pp. 1--9.

\end{thebibliography}


\end{document}